# Anisotropic magnetoresistance in antiferromagnetic semiconductor $Sr_2IrO_4$ epitaxial heterostructure

X. Marti[1,2,3,*], I. Fina[4], D. Yi[1], J. Liu[5], J.H. Chu[5], C. Rayan-Serrao[1], S. Suresha[6], J. Železný[3], T. Jungwirth[3,7], J. Fontcuberta[4], R. Ramesh[1,5,6,8]

[1]Department of Materials Science and Engineering, University of California, Berkeley, California 94720, USA

[2]Department of Condensed Matter Physics, Faculty of Mathematics and Physics, Charles University, 12116 Praha 2, Czech Republic

[3]Institute of Physics ASCR, v.v.i., Cukrovarnická 10, 162 53 Praha 6, Czech Republic

[4]Institut de Ciència de Materials de Barcelona, ICMAB-CSIC, Campus UAB, E-08193 Bellaterra, Spain

[5]Department of Physics, University of California, Berkeley, California 94720, USA

[6]National Center for Electron Microscopy, Lawrence Berkeley National Laboratory, Berkeley, California 94720, USA

[7]School of Physics and Astronomy, University of Nottingham, Nottingham NG7 2RD, United Kingdom

[8]Materials Science Division, Lawrence Berkeley National Laboratory, Berkeley, California 94720, USA

* xaviermarti@berkeley.edu

Lord Kelvin with his discovery of the anisotropic magnetoresistance (AMR) phenomenon[1] in Ni and Fe was 70 years ahead of the formulation of relativistic quantum mechanics the effect stems from, and almost one and a half century ahead of spintronics whose first commercial applications relied on the AMR.[2] Despite the long history and importance in magnetic sensing and memory technologies, the microscopic understanding of the AMR has struggled to go far beyond the basic notion of a relativistic magnetotransport phenomenon arising from combined effects on diffusing carriers of spin-orbit coupling and broken symmetry of a metallic ferromagnet. Our work demonstrates that even this seemingly generic notion of the AMR phenomenon needs revisiting as we observe the ohmic AMR effect in a nano-scale film of an antiferromagnetic (AFM) semiconductor $Sr_2IrO_4$ (SIO).[3] Our work opens the recently proposed[4,5] path for integrating semiconducting and spintronic technologies in AFMs. SIO is a particularly favorable material for exploring this path since its semiconducting nature is entangled with the AFM order and strong spin-orbit coupling.[3,6,7] For the observation of the low-field Ohmic AMR in SIO we prepared an epitaxial heterostructure comprising a nano-scale SIO film[8,9] on top of an epilayer of an FM metal $La_{2/3}Sr_{1/3}MnO_3$ (LSMO). This allows the magnetic field control of the orientation of AFM spins in SIO via the exchange spring effect at the FM-AFM interface.[10,11,12]

Our work emerges from the combination of three recently proposed areas of research in relativistic physics phenomena, nanostructure fabrication, and spintronics technologies. The first one is represented by large and bistable magnetoresistance signals that have been observed in tunneling devices with a nano-scale AFM metal film of IrMn on one side and a non-magnetic metal on the other side of the tunnel barrier.[10,11,12] The work has experimentally demonstrated the feasibility of a spintronic concept in which the electronic device characteristics are governed by an AFM. It has been shown that the thin-film AFM moments can be manipulated via an exchange coupled FM[10,12] and that the AFM magnetoresistance effect can persist in these devices to room temperature.[13] Without the auxiliary FM, AFM-based spintronics elements may offer unprecedented opportunities for dense and robust non-volatile storage of information because of the lack of magnetic stray fields of compensated-moment AFMs and by utilizing the relative insensitivity of the materials in the AFM ordered state to external magnetic fields.[14,15]

The second root is based on the recognition that there are fundamental physical limitations for metallic magnets which may make them impractical to realize the full potential of spintronics. In particular, metals are unsuitable for transistor and information processing applications or for photonics. The synthesis of semiconductors with robust FM ordering of spins, which would simultaneously enable the conventional tunability of electronic properties, remains a significant challenge.[16] AFM ordering occurs much more frequently in nature than FM ordering, particularly in conjunction with semiconducting behavior.[17,18] Recently, it has been demonstrated that high Néel temperature AFM counterparts of the conventional compound semiconductors can be prepared in thin epitaxial films and heterostructures, opening the prospect for AFM semiconductor based microdevices.[4,5,19]

The third root of our work is the recent discovery of the unconventional Mott insulating state induced by the relativistic spin-orbit coupling and AFM order in SIO,[3,6,7] and the demonstration of tunable semiconducting properties of thin epitaxial SIO films.[8,9] The entangled semiconducting electronic structure and magnetic ordering suggest that electron transport and magnetic state might be strongly coupled in SIO. This is reminiscent of the excitement that has driven the research field of FM semiconductors in which such a coupling originates from the carrier-mediated nature of the ferromagnetic order.[16] Finally, the strong spin-orbit coupling ties SIO with the proposed concept of spintronics based on the relativistic anisotropic magnetotransport phenomena. The common characteristics of these effects are that they are an even function of the microscopic magnetic moment vector which makes them equally well present in spin-orbit coupled AFMs as in FMs.[20]

We have previously prepared high-quality SIO epilayers on $SrTiO_3$ (STO) substrates using pulsed laser deposition assisted by reflection high-energy electron diffraction (RHEED).[8] The SIO films show activated, temperature dependent transport corresponding to a semiconductor with a band gap of 10's to 100's meV, depending on the lattice mismatch-induced strain in the SIO epilayer.[8,9] Apart from temperature and epitaxial strain, strong variations in the resistivity of these SIO films can be induced by doping or electrostatic gating.[21] In this work we show that SIO displays the spin-orbit coupling induced AMR, opening the route towards the development of spintronic devices in the remarkable family of these AFM Mott semiconductors. We report sizable AMR signals in a geometry in which the angle of the AFM moments is fixed with respect to the current direction and varies only relative to the crystal axes. In conventional FM metals, this geometry yields typically weak AMR signals which points to new physics and device concepts that the oxide AFM semiconductors may introduce into the field of spintronics.

To allow for the rotation of the SIO AFM moments via the FM-AFM exchange spring effect, we have prepared a heterostructure comprising an epitaxial 12 nm thick film of a FM-metal LSMO inserted between a (001)STO single crystalline substrate and a 6 nm thick film of the AFM-semiconductor SIO. The high epitaxial quality of our materials is illustrated in Fig. 1a which shows a scanning transmission electron microscopy image of the SIO and LSMO heterostructure. Atomic force microscopy of the top SIO surface indicates that the multilayer preserved the substrate root-mean-square roughness below one unit cell. Both LSMO and SIO grow **c**-axis oriented ([001]STO//[001]LSMO//[001]SIO) with an in-plane epitaxial relationship [100]STO//[100]LSMO//[110]SIO. (See Supplementary Information for further growth and structural details.) In Fig. 1c we plot the temperature dependent magnetization of our SIO/LSMO stack at 100 mT magnetic field **B**//[100]. The figure demonstrates the FM transition of LSMO above room temperature and a cusp of the magnetic susceptibility at $T^* \sim 100$ K. The applied magnetic field strength is over one order of magnitude larger than the coercive field observed in magnetization loops collected at T = 4.2 K (see Supplementary Information). The detected splitting of the zero-field-cooled and field-cooled magnetization curves below T* is consistent with the canted AFM ground state of SIO.[3] The Néel temperature $T_N \sim T^*$ is smaller than the bulk value (240 K) which is expected for an ultrathin AFM film, in our case with less than three unit cells along the **c**-axis.

In Fig. 1a we depict the electrical measurement geometries used in our transport experiments. In the first geometry, one electrode (1) is attached on the free SIO surface and electrical current **I** is driven vertically through SIO and then laterally through LSMO along the [100] axis between the contact (1) and the side contact (4). We also performed control measurements in a lateral geometry in which current is driven only laterally between the side contacts (4) and (5) and voltage is measured between additional lateral contacts (2) and (3) attached to LSMO (see Supplementary Information for further details).

In Fig. 1d we compare the temperature-dependent transport characteristics of our device measured in the lateral LSMO geometry and in the vertical SIO – lateral LSMO geometry. While in both cases the current-voltage (I-V) characteristics remain Ohmic in the entire temperature range from room temperature to 4.2 K (4.2 K data are shown in the inset of Fig. 1d), the temperature dependence of the resistivity is vastly different in the two configurations. The data confirm metallic conduction through LSMO and semiconducting transport in SIO.

The central result of our work is shown in Fig. 1b. At sufficiently low temperatures (4.2 K in Fig. 1b), we apply an external in-plane magnetic field of amplitude 350 mT at angles $\theta_1 = 45°$ and $\theta_2 = 135°$ measured from the [100] axis, as defined in Fig. 1a. Note that the field is much smaller than typical AFM exchange fields in AFM oxides with super-exchange induced magnetic order[17] and larger than the LSMO coercivity which is approximately 5 mT at 4.2 K. For the two angles $\theta_1$ and $\theta_2$, we observe two distinct and switchable resistance states when current is driven through the SIO/LSMO stack and temperature is sufficiently low so that the vertical transport path across the SIO semiconductor-AFM dominates the resistance. In the following, we discuss in detail the phenomenology and the origin of the detected AMR in our AFM semiconductor.

In Figs. 2a-f we compare field-rotation AMR data at B =100 mT as a function of temperature detected in the two measurement configurations (see sketches in Fig. 2). At higher temperature and for the lateral LSMO transport geometry (Fig. 2d), we observe the typical Ohmic AMR = [R(θ) - R(0)]/R(0) of a FM material.[22] In a rotating in-plane magnetic field of strength larger than the coercive field, the FM magnetic moment **m** follows **B**. In the lateral transport geometry this implies that the in-plane angle θ between **m** and the electrical current **I**//[100] vary which results in the observed AMR signal proportional to sin(2θ). The observation that the resistance for **I** // **m** is smaller than for **I** ⊥ **m** is in agreement with previous reports on AMR in LSMO.[23] Since at low temperature LSMO is a metal with a weak temperature dependence of its resistivity, the lateral AMR signal of LSMO is also only weakly temperature dependent (Figs. 2d-e). The observed small departure from the smooth cos(2θ) angular dependence at low temperatures is caused by an enhanced coercivity of LSMO when decreasing temperature (see Supplementary Information).

Measurements for the current driven through the SIO/LSMO stack are shown in Figs. 2a-c. At high temperature (200K in Fig. 2a) when the resistance of the transport channel is dominated by the long lateral path in LSMO, we observe an AMR $\sim \sin(2\theta)$ as in Fig. 2d. The amplitude of the AMR is partly suppressed due to the contribution to the measured resistance from a signal arising from the transport path in SIO which is non-magnetic at 200 K. Below 100 K (Fig. 2b) the resistance of the short vertical path through SIO becomes dominant and the AMR is diminished. Remarkably, when sufficiently below the Néel temperature of SIO (see Fig. 2c with 4.2 K data), the AMR signal increases again. Since the resistance at these low temperatures is completely dominated by the vertical transport path in SIO, the observed Ohmic AMR signal originates from the semiconductor-AFM layer.

A clear signature of the distinct origin of the low-temperature AMR is the higher harmonic component of the SIO AMR signal seen in Fig. 2c. To highlight this point we compare in Fig. 3a,b polar plots of the SIO and LSMO AMRs measured at 4.2 K and 350 mT. In Figs. 3c,d we show complete resistance maps which were all collected after field-cooling the sample from 300 K to 4.2 K in a 350 mT magnetic field applied along the in-plane angle $\theta$ = 90°. With the temperature maintained at 4.2 K, subsequent field-rotation AMR measurements were performed at magnetic fields of strength varying from 350 mT to 5 mT. The resistance maps in Figs. 3c,d confirm the distinct phenomenology of the SIO and LSMO governed AMR signals.

Our experiments indicate that the observed oscillatory AMR response of SIO does not originate from a direct action of the weak applied magnetic field on the magnetic ordering of the AFM semiconductor. (For the discussion of the direct field effects observed previously in bulk SIO[24] see Supplementary information.) It is rather an indirect consequence of rotating moments in the exchange-coupled FM LSMO layer which drag the AFM moments in SIO. We point out that we designed our stack with the relatively small thickness of SIO compared to LSMO to maximize the exchange-spring effect of the FM on the AFM.[10,12,13,25] Fig. 4 provides a key evidence for this scenario. We have prepared a control epitaxial stack in which a non-magnetic LaNiO$_3$ (LNO) metal film of thickness 4 nm is inserted between the SIO and LSMO to break the FM-AFM interlayer exchange coupling. The crystal-quality (Fig. 4b), magnetization (see Supplementary Information) and resistivities are comparable in the SIO/LNO/LSMO (Fig. 4c) and SIO/LSMO (Fig. 1d) stacks. However, the low-temperature oscillatory AMR signal in the vertical transport geometry (Fig. 4d) is absent in the control SIO/LNO/LSMO sample with uncoupled AFM and FM layers.

The exchange-coupling between LSMO and SIO in the SIO/LSMO stack is also evidenced by a strong increase of the coercivity of LSMO at $T < T_N$. Note that the marked broadening of the hysteresis loop is absent in bare LSMO films or in the SIO/LNO/LSMO control structure (see Supplementary Information). The strong enhancement of the coercivity of LSMO in the SIO/LSMO stack, therefore, provides another evidence for the exchange-spring induced rotation of AFM moments in our AMR measurements in SIO.

In Figs. 5a,b we compare the magnetoresistance at 4.2 K for the lateral LSMO and vertical SIO – lateral LSMO transport measurements performed after field-cooling the sample in magnetic field applied at the angle $\theta = 90º$ (red curves). In both transport geometries, the $\Delta R(B)/R(0)$ data display closely related features, confirming that the SIO layer response follows the magnetization of LSMO. The coupling between the AFM and FM layers becomes even more evident if the magnetoresistance is subsequently measured after aligning the field at $\theta = 180º$ (see black curves in Figs. 5a,b). Here the lateral LSMO transport measurement shows a pronounced hysteretic double peak which signals an abrupt magnetic domain reconfiguration. The observation that $\Delta R(B)/R(0)$ of LSMO differs when performing the field sweeps along the field-cooling direction or perpendicular to it reflects the existence of a pinned magnetic domain configuration. Remarkably, this domain reconfiguration is reflected in the vertical SIO - lateral LSMO measurement, where the much larger resistance of SIO dominates the overall resistance. It confirms that the magnetic structure of the SIO layer responds to that of the underlying LSMO layer, and highlights the exchange-spring magnetic coupling between LSMO and SIO. We also point out that, consistently, the magnitude of $\Delta R(B)/R(0)$ in SIO is comparable to the magnitude of the corresponding field rotation AMR signal in SIO (compare Fig. 5 and Fig. 3). Similar experiments to those reported in Fig. 5a,b have been performed over the entire range of angles $0° < \theta < 360°$. The resulting SIO magnetoresistance sweep map, shown in Fig. 5c. has common as well as some distinct features compared to the SIO rotation AMR map in Fig. 3c which are due to the presence of hysteretic effects within the measured field range.

We point out that the AMR we observe in our semiconductor-AFM is of the order of 1 %. This is a relatively large signal for an Ohmic magnetic device. Moreover, SIO is an easy-plane AFM3 while the electrical current is driven through SIO along the out-of-plane **c**-axis. This implies that for the in-plane field rotations the "non-crystalline" contribution to the AMR, i.e. the AMR corresponding to the varying angle between magnetic moments and current, does not contribute to our signal in SIO. We are detecting only the typically weaker "crystalline" AMR due to magnetic moment rotation relative to the crystal axes.[26] Still the AMR signal in SIO is comparable to typical non-crystalline Ohmic AMRs in FMs. We also point out that the sizable AMR signal is measured in our proof-of-concept device with a simple geometry and macroscopic contacts which underlines the potential of semiconductor-AFMs for realizing the goal of integrated semiconducting and spintronics functionalities in one material.

The observation of the Ohmic AMR in SIO opens intriguing microscopic physics questions on its origin. Common theories of the Ohmic AMR in magnetic materials associate the phenomenon with anisotropic scattering induced by the spin-orbit coupling.[2,22] This applies to both the dependence on magnetic moment orientation with respect to current or to crystal axes, with the latter reflecting the crystal symmetries in the angular dependence of the AMR. The spin-orbit coupling induced anisotropic scattering, combined with the in-plane cubic anisotropy of the SIO lattice, can be one origin of our observed Ohmic AMR. However, the entanglement of spin-orbit coupling and magnetic order with the semiconducting electronic structure offers also a complementary scenario in which the magnetic moment rotations change the band gap, and consequently the density of mobile carriers and the resistance. The canted AFM ground state of SIO adds to the complexity of this scenario but it also underlines its potential importance. Recent studies have indicated that the semiconductor band gap is indeed sensitive to the bonding and the AFM canting angles which are mutually coupled by the Dzyaloshinskii-Moriya exchange interaction.[27,36,8]

To conclude, our observation of the Ohmic AMR in SIO thin-films of amplitude comparable to AMR in common metallic FMs demonstrates that the AMR phenomenon can be observed in a broad class of systems beyond those with an FM order, and confirms the viability of the recently proposed concept of AFM semiconductor spintronics. Since many AFM semiconductors, unlike FM semiconductors, have high ordering temperatures, our work may open the path towards semiconductor spintronics devices operating at room temperature. The semiconducting electronic structure of SIO is intimately related to the AFM order, spin-orbit coupling and electron correlations. This implies that unprecedented microscopic physics mechanism may play an important role in the relativistic AFM spintronic effects in SIO films and nano-devices.

**Acknowledgements**

The authors acknowledge the support from the NSF (Nanosystems Engineering Research Center for Translational Applications of Nanoscale Multiferroic Systems, Cooperative Agreement Award EEC-1160504) and DOE. T.J. acknowledges ERC Advanced Grant 268066 and Praemium Academiae of the Academy of Sciences of the Czech Republic. X.M. acknowledges the Grant Agency of the Czech Republic No. P204/11/P339. Financial support by the Spanish Government (Projects MAT2011-29269-C03, CSD2007-00041, MAT2010-16407 and CSD2009-00013), Generalitat de Catalunya (2009 SGR 00376) is acknowledged.

**Author contributions**

Device fabrications, D.Y., J.L., C.R.-S.; scanning transmission electron microscopy, S.S.; experiments and data analysis, I.F., X.M., T.J., J.F., J.H.C., J.L., J.Ž, D.Y., R.R.; writing and project planning, X.M., T.J., J.F., R.R.

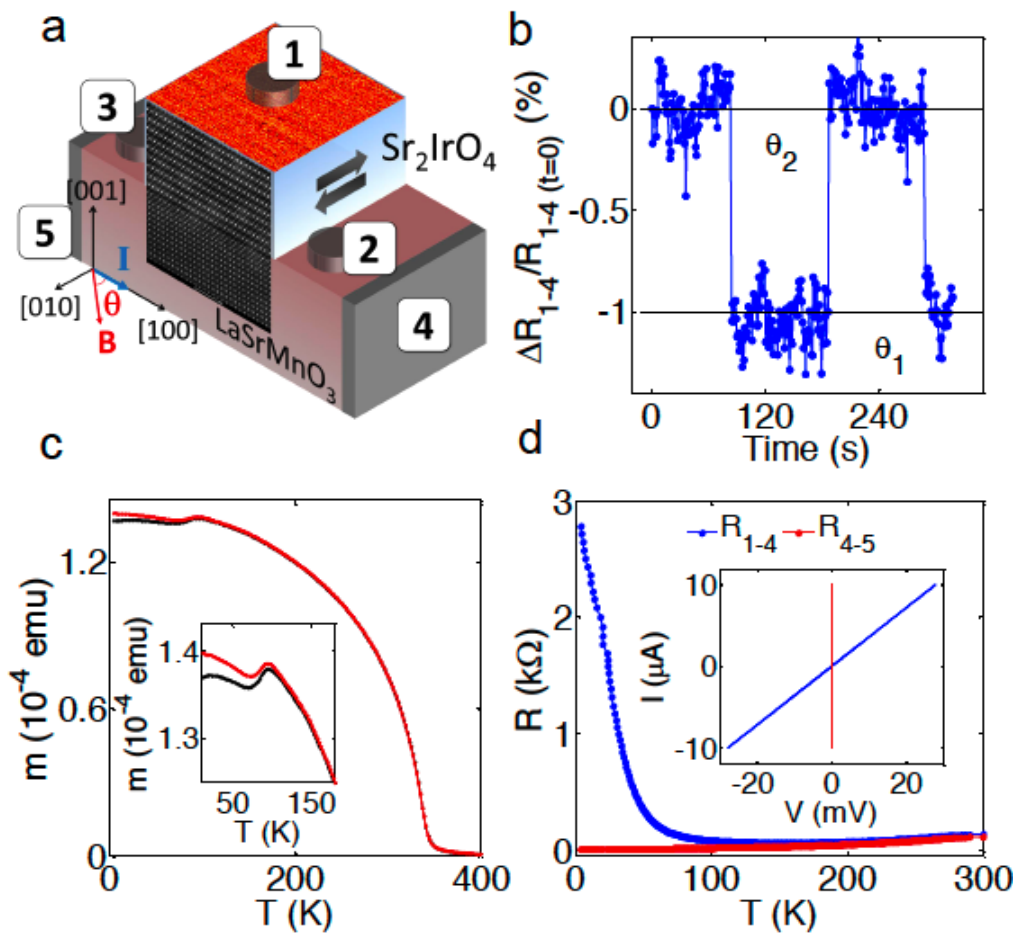

**Figure 1** | **a**, Sketch of the transport device and transmission electron micrograph of the heterostructure. SIO (blue) is epitaxially grown on top of a LSMO film (red). The high quality of the materials is illustrated by the scanning transmission electron micrograph and the surface topography image whose larger versions are in the Supplementary Information. Labels enumerate the electrical contacts of the device. **b**, Stable resistance states measured between contacts (1) and (3) at applied magnetic field angles alternating between $\theta_1 = 45°$ and $\theta_2 = 135°$. **c**, Zero field-cooling (black) field-cooling (red) magnetization of the heterostructure for B of 100 mT applied along the [010] axis showing the LSMO FM transition at $T_C$ and the onset of the canted AFM ordering in SIO at $T_N$. **d**, Resistance as a function of temperature measured between contacts (1), (4) and (4), (5    ) showing the semiconducting SIO and metallic LSMO, respectively. The inset shows corresponding Ohmic I-V characteristics measured at 4.2 K.

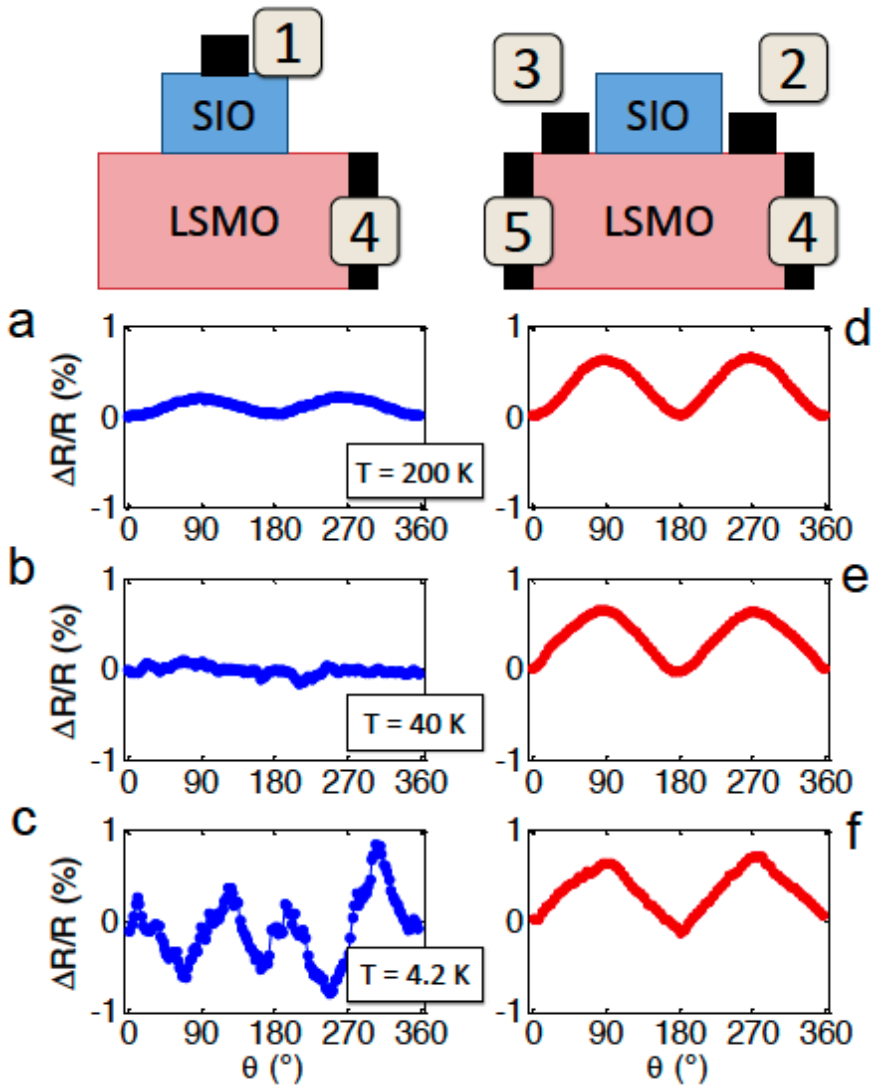

**Figure 2 |** AMR as a function of the temperature for the two sketched configurations (top) measured after applying 100 mT in-plane rotating magnetic field. **a,b,c,** In the vertical geometry (left), we observe three distinct temperature regimes of the device operation: at high temperatures the AMR is dominated by the metal FM LSMO, at intermediate temperatures no AMR is observed, and at sufficiently low temperatures the observed AMR signal is due to the AFM semiconductor SIO. **d,e,f,** In the control lateral measurements the LSMO AMR is probed. At higher temperatures, the $\sin(2\theta)$ AMR is observed, at intermediate temperatures, below $T_N$, the coupling to SIO produces a marked departure from the $\sin(2\theta)$ dependence due to the enhanced LSMO coercivity.

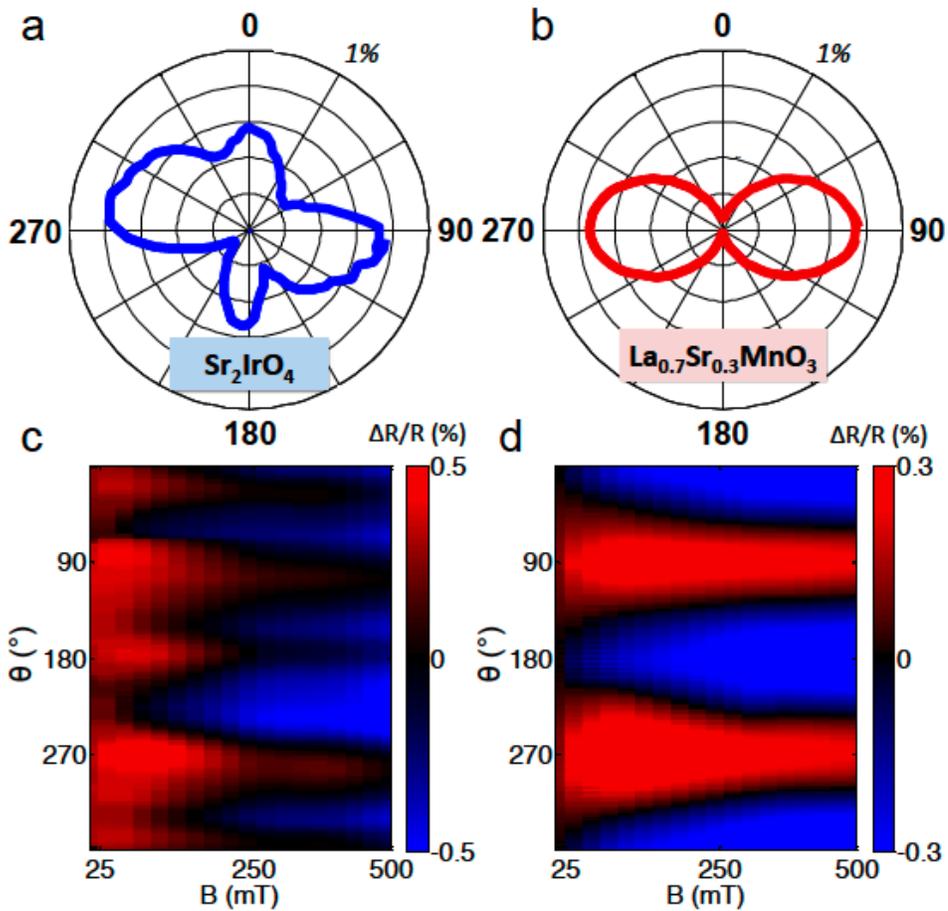

**Figure 3 | a,b** Polar plots corresponding to AMR experiments in a 350 mT in-plane rotating field and temperature 4.2 K. The LSMO AMR shows the $\sin(2\theta)$ behavior at this large magnetic field. The SIO AMR has a distinct, higher harmonic component. **c,d** Field-rotation AMR maps demonstrating the coupling of the LSMO and SIO moments.

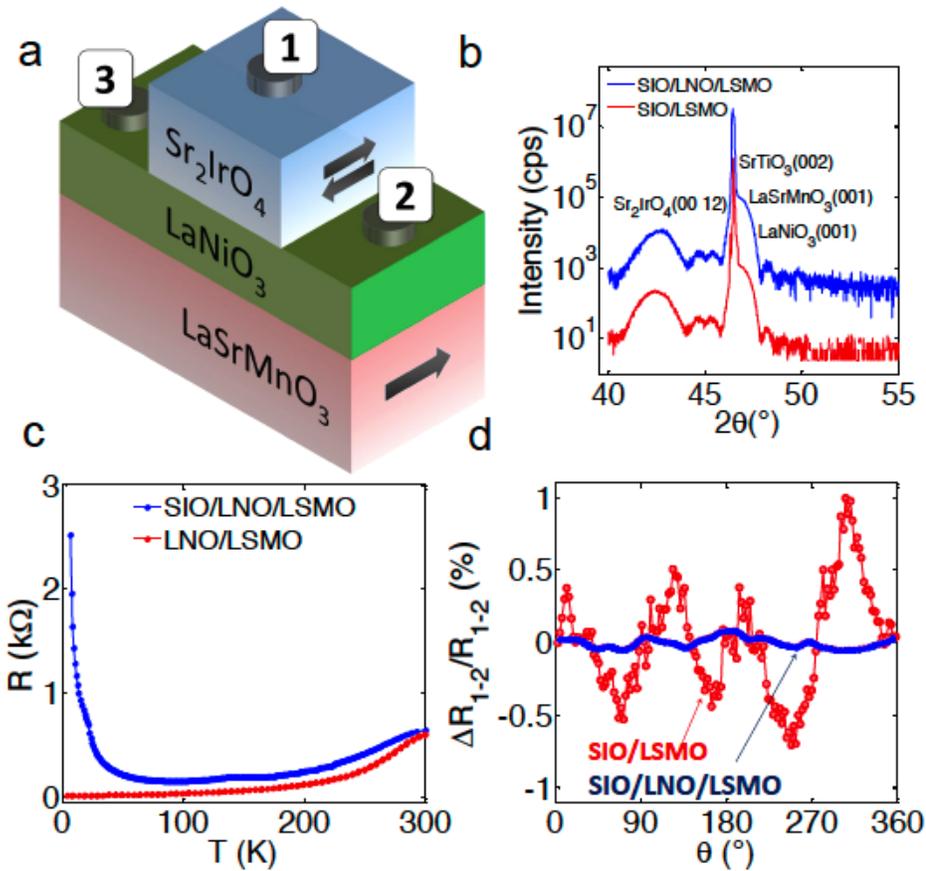

**Figure 4 | a**, Sketch of the control sample in which a 10 unit cell (4nm) paramagnetic metal LNO is inserted between the LSMO and SIO layers in order to break the FM-AFM coupling. **b**, X-ray diffraction patterns of the SIO/LSMO and SIO/LNO/LSMO samples confirming the presence of the SIO in the control SIO/LNO/LSMO stack with the same layer thickness and crystal quality as in the SIO/LSMO sample. **c**, Electrical resistance measured in the control SIO/LNO/LSMO sample indicating that LSMO is metallic and that the SIO layer has comparable electrical resistance and the same semiconducting character as in the SIO/LSMO stack. **d**, Resistance measurement in an in-plane rotating magnetic field of 100 mT in the SIO/LNO/LSMO sample compared to the AMR data in the SIO/LSMO stack. The SIO AMR is not observed in the control SIO/LNO/LSMO sample at 100 mT as well as at higher applied fields.

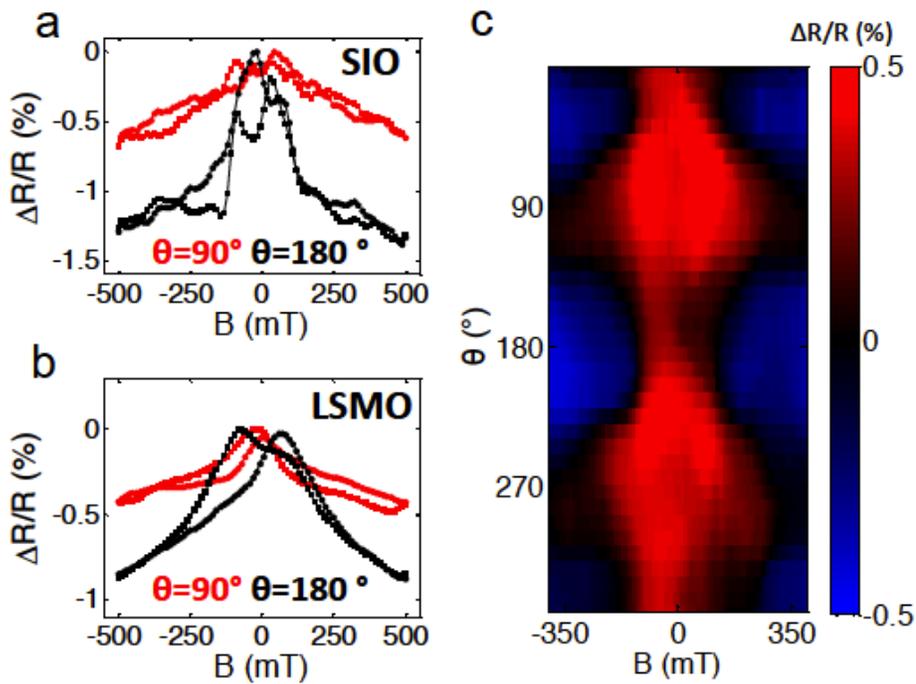

**Figure 5 | a**, Magnetoresistance of SIO after field-cooling to 4.2 K in applied magnetic field along θ = 90° and subsequently sweeping the magnetic field at the same magnetic field angle and at θ = 180°. **b,** Same as **a,** for the lateral LSMO magnetoresistance. **c**, Field-sweep magnetoresistance map of SIO for the entire range of field angles. All measurements in were performed after the field-cooling along θ = 90°.